\documentclass[10pt,conference]{IEEEtran}
\IEEEoverridecommandlockouts
\usepackage{cite}
\usepackage{amsmath,amssymb,amsfonts}
\usepackage{algorithmic}
\usepackage{graphicx}
\usepackage{textcomp}
\usepackage{xcolor}
\usepackage[utf8]{inputenc}
\usepackage[T1]{fontenc}
\usepackage[hidelinks]{hyperref}

\usepackage{caption}
\usepackage{subcaption}

\usepackage{url}

\usepackage{cleveref}
\def\BibTeX{{\rm B\kern-.05em{\sc i\kern-.025em b}\kern-.08em
    T\kern-.1667em\lower.7ex\hbox{E}\kern-.125emX}}
\begin{document}

\title{Developing Apps for Researching the COVID-19 Pandemic with the TrackYourHealth Platform\\
{\footnotesize
}
}%
\ifx\doubleBlind\undefined
\author{
	\IEEEauthorblockN{
		Carsten Vogel\IEEEauthorrefmark{1},
		Rüdiger Pryss\IEEEauthorrefmark{1},
		Johannes Schobel\IEEEauthorrefmark{2},
		Winfried Schlee\IEEEauthorrefmark{3}
		and
		Felix Beierle\IEEEauthorrefmark{1}
	}
	\IEEEauthorblockA{
		\IEEEauthorrefmark{1}Institute of Clinical Epidemiology and Biometry,
		University of Würzburg,
		Würzburg, Germany\\
		Email: \{carsten.vogel, ruediger.pryss\}@uni-wuerzburg.de, beierle@tu-berlin.de
	}
	\IEEEauthorblockA{\IEEEauthorrefmark{2}DigiHealth Institute,
		Neu-Ulm University of Applied Sciences,
		Neu-Ulm, Germany\\
		Email: johannes.schobel@hnu.de
	}
	\IEEEauthorblockA{\IEEEauthorrefmark{3}Clinic and Policlinic for Psychiatry and Psychotherapy,
		University of Regensburg,
		Regensburg, Germany\\
		Email: winfried.schlee@ieee.org
	}
}
\else
\author{
	\IEEEauthorblockN{
		\anonymize{xxxxxxxxxxxxxxxx}\IEEEauthorrefmark{1},
		\anonymize{xxxxxxxxxxxxxxxx}\IEEEauthorrefmark{1},
		\anonymize{xxxxxxxxxxxxxxxx}\IEEEauthorrefmark{2},
		\anonymize{xxxxxxxxxxxxxxxx}\IEEEauthorrefmark{3}
		and
		\anonymize{xxxxxxxxxxxxxxxx}\IEEEauthorrefmark{1}
	}
	\IEEEauthorblockA{\IEEEauthorrefmark{1}\anonymize{xxxxxxxx,}
		\anonymize{xxxxxxxx},
		\anonymize{xxxxxxxx}, \anonymize{xxxxxxxx}\\
		Email: \{\anonymize{xxxxxxxxxxxx}, \anonymize{xxxxxxxxxxxx}, \anonymize{xxxxxxxxxxxx\}}@\anonymize{xxxxxxxxxxxx}
	}
	\IEEEauthorblockA{\IEEEauthorrefmark{2}\anonymize{xxxxxxxx,}
		\anonymize{xxxxxxxx},
		\anonymize{xxxxxxxx}, \anonymize{xxxxxxxx}\\
		Email: \anonymize{xxxxxxxxxxxx}@\anonymize{xxxxxxxxxxxx}
	}
	\IEEEauthorblockA{\IEEEauthorrefmark{3}\anonymize{xxxxxxxx,}
		\anonymize{xxxxxxxx},
		\anonymize{xxxxxxxx}, \anonymize{xxxxxxxx}\\
		Email: \anonymize{xxxxxxxxxxxx}@\anonymize{xxxxxxxxxxxx}
	}
}
\fi
\maketitle
\begin{abstract}
Through lockdowns and other severe changes to daily life, almost everyone is affected by the COVID-19 pandemic.
Scientists and medical doctors are -- among others -- mainly interested in researching, monitoring, and improving physical and mental health of the general population.
Mobile health apps (mHealth), and apps conducting ecological momentary assessments (EMA)
respectively, can help in this context.
However, developing such mobile applications poses many challenges like costly software development efforts, strict privacy rules, compliance with ethical guidelines, local laws, and regulations.
In this paper, we present TrackYourHealth (TYH), a highly configurable, generic, and modular mobile data collection and EMA platform, which enabled us to develop and release two mobile multi-platform applications related to COVID-19 in just a few weeks.
We present TYH and highlight specific challenges researchers and developers of similar apps may also face, especially when developing apps related to the medical field.
\end{abstract}

\begin{IEEEkeywords}
COVID-19,
Mobile Application Development,
Ecological Momentary Assessment,
Mobile Crowdsensing
\end{IEEEkeywords}

\section{Introduction}
\label{sec:introduction}
In early 2020, the coronavirus spread all around the globe and posed major challenges to all areas of society.
Not only does COVID-19 affect the respiratory system of infected people, it also impacts the social consequences and measures trying to contain the virus.
Lockdowns, home office, restricted everyday routines, limited freedom of movement and social distancing take their toll on everyone.
Studies have found that overall mental health has declined \cite{
	pieh2020EffectAgeGenderIncomeWork
	,vindegaard2020COVID19PandemicMentalHealthConsequences
}.

In order to find out about the mental and physical health of people during sudden events like the COVID-19 pandemic,
mobile applications with ecological momentary assessment (EMA) and mobile crowdsensing (MCS) features are a viable solution.
Employing mobile applications has two major benefits for both researchers and potentially also its users:
(1) advancing research
and
(2) supporting people to cope.
However, in most cases, researchers demand such an app that can be easily adapted
to their needs, such as the changing COVID-19 situation might require.
Besides developing the software itself, researchers face additional challenges
that are costly and time consuming, like compliance with
privacy regulations, ethical guidelines,
local laws, and platform regulations.
At the same time, while facing these challenges, unforeseen events like the current pandemic make a short time-to-market crucial.

In this paper, we present TrackYourHealth (TYH), a modular server-client platform for mobile apps.
TYH's core features constitute EMA -- users can fill out questionnaires, data is collected, and individualized feedback can be sent back to the user from the backend.
We present
specific challenges researchers and developers
face when developing apps related to critical health-related
situations like the current pandemic.
Our findings can help gauge the effort necessary for each step.

Based on TYH, within a matter of weeks, we developed and released two multi-platform COVID-19-related mobile apps, which were used by 131,371 users at the time of writing.
\ifx\doubleBlind\undefined
The first one is Corona Check\footnote{CC: \url{https://www.coronacheck.science/en/}; last accessed: 2021-02-22} (CC), developed in collaboration with a regional health ministry in Germany.
\else
The first one is \anonymize{xxxx xxxx xxx}\footnote{CC: \anonymize{xxxxxxxxxxxxxxxxxxxxxxxxxxxxxx}; last accessed: 2021-02-22} (CC), developed in collaboration with a regional health ministry in Germany.
\fi
Users can enter their symptoms to get a first impression if they could be linked to the coronavirus, novel at that time.
The main goal of CC was to relief the corona telephone hotlines, to reduce uncertainty among the population by providing reliable information, and to collect data for further research.
\ifx\doubleBlind\undefined
The Corona Health\footnote{CH: \url{https://www.corona-health.net/en/}; last accessed: 2021-02-22} (CH) app was developed in cooperation with the Robert Koch Institute (RKI), Germany's governmental institute responsible for disease control and prevention.
\else
The \anonymize{xxxx xxxxxxx}\footnote{CH: \anonymize{xxxxxxxxxxxxxxxxxxxxxxxxxxx}; last accessed: 2021-02-22} (CH) app was developed in cooperation with the Robert Koch Institute (RKI), Germany's governmental institute responsible for disease control and prevention.
\fi
The main goal of CH is to conduct several studies on the direct and indirect effects of the COVID-19 virus and its countermeasures on the mental and physical health of the population.

In the following, we present the TYH platform that allowed us to implement these apps in a matter of weeks.
In Section \ref{sec:relatedWork}, we discuss related work.
In Section \ref{sec:configurableMobileApplicationFramework},
we elaborate TYH's core features and user privacy aspects.
In Section \ref{sec:timeline}, we present the timeline for CC and CH,
from the basic idea to users being able to find the final product in the app store.
In Section \ref{sec:conclusionAndFutureWork},
we conclude and give an outlook on future work.

\section{Related Work}
\label{sec:relatedWork}
COVID-19-related tracking apps raise concerns among users, data protection authorities, and researchers regarding user data, security, and privacy \cite{
	ahmed2020SurveyCOVID19ContactTracingApps
	,borra2020COVID19AppsPrivacySecurityConcerns
}.
We have approached this with transparency and strict compliance to all relevant regulations.
\cite{beierle2018TYDRTrackYourDailyRoutine} presents an app with elaborated concepts
for MCS and EMA.
With \cite{vandeven2017ULTEMATMobileFrameworkSmartEcological}, a framework was introduced for
time-based scheduling of EMAs and
interventions.
Both are not implemented as a multi-platform and are only available for Android.

There are some existing multi-platform frameworks for EMA and MCS like
AWARE\footnote{AWARE: \url{https://awareframework.com}; last accessed: 2021-02-22}, which enable sensor data tracking and offer studies with questionnaires.
However, building upon such frameworks would likely have lead us to make several adjustments to adapt to the specific requirements of our use cases.
Additionally, compliance to regulations regarding medical products
required us to know and describe every
feature of the platform.
\ifx\doubleBlind\undefined
For CC and CH, we built on the TYH platform, which we previously used to
 create other EMA apps like TrackYourTinnitus \cite{probst2016EmotionalStatesMediatorsTinnitusLoudness} or TrackYourStress \cite{pryss2019ExploringTimeTrendStressLevels}.
\else
For CC and CH, we built on the TYH platform, which we previously used to create other EMA apps like \anonymize{xxxxxxxxxxxxxxx} \cite{blindedSource1} or 
\anonymize{xxxxxxxxxxxxxx} \cite{blindedSource2}.
\fi

\section{The Track Your Health Platform}
\label{sec:configurableMobileApplicationFramework}

In this section, we present the TYH platform and focus on overall requirements, core features and user privacy aspects.

\subsection{Overall Requirements}
\label{sec:frameworkRequirements}

General requirements for EMA apps are that questionnaires
with different input types are available in order to conduct
research.
Additionally, we conduct MCS.
Every time a users fills out a questionnaire,
data from mobile sensors of the device,
e.g., the user's location or app usage statistics,
can help support researching behavior and health conditions.
Especially with respect to mobile sensing, data protection requirements
and regulations have to be complied with.
Complementary to data collection, for our applications, we also required a feedback channel that responds to the user input.
In the case of CC and CH,
we had additional specific requirements
regarding the delivery of corona-related news to the user,
how feedback from questionnaires is presented, etc.
Further requirements in our case were that the apps should be multilingual and that the platform has to react to changes during live operation.

\subsection{Architecture}
\label{sec:architecture}
TYH comprises a client-server-architecture with native mobile applications (Android and iOS), communicating JSON data structures via HTTPS to a REST API.
All structural data and contents are JSON formatted objects provided by the server, stored by the mobile applications locally once delivered and refreshed if necessary.
Likewise, the server itself as well as the mobile applications are configured via JSON files.

\subsection{Server Features}
\label{sec:serverFeatures}
\ifx\doubleBlind\undefined
The TYH server API \cite{pryss2018RequirementsFlexibleGenericAPIEnabling} is developed in PHP by utilizing the Laravel framework and follows a RESTful approach.
\else
The TYH server API \cite{blindedSource3} is developed in PHP by utilizing the Laravel framework and follows a RESTful approach.
\fi
Furthermore, a relational database (i.e., MySQL) is used as data storage.
The API itself comprises various modules, accessible through well-defined and documented endpoints.
The API covers features like authentication and authorization of users, provides details for currently running studies as well as assigned questionnaires.
In addition, users can send data captured using their smart mobile devices to the server, that is then appropriately validated to ensure consistently high data quality.
Furthermore, it is possible to define automatically generated rule-based feedback that can be requested by the mobile clients after a questionnaire has been sent.
The API can also host a messaging system that allows users to communicate with their responsible healthcare provider (i.e., psychologist or medical doctor).
Note that the one-on-one contact to healthcare providers was not integrated for CC and CH
because of the high number of users we expected.
The client-server data exchange strictly follows the JSON:API specification\footnote{JSON:API: \url{https://jsonapi.org}; last accessed: 2021-02-22}.
Caching, in turn, is utilized based on shallow \emph{ETags} generated by the server for every response. 

\subsection{Client Features}
\label{sec:clientFeatures}
The mobile apps are developed in Java for Android 5+ and with Swift for iOS 12+.
They consist of several modules that work independently.
Each module is intended for a specific functionality such as filling out questionnaires, giving feedback to the user, or managing user accounts, etc.
Depending on the use case, subsets of modules can be included into the application and configured as desired.
In general, almost everything, from the UI theme and icons over the subset, order, and settings of modules to the content and translations, can be customized via configuration files.
While the theme and base layout are delivered within the application itself, some dynamic content, like questionnaires, feedback, studies, news etc. is loaded over the TYH API.
This concept enables updates during live operation.

The central module is the \emph{User-Account}, which provides basic authentication functionality like registration via email or username and password or anonymously.
The module handles email validation, the login itself, providing a token and user profile as well as various settings used by other modules.
All requests that require authentication are token-based.

The client framework provides a module that enables users to leave and join available \emph{Studies} as seen in Figure \ref{fig:corona-health-stress-study}.
This, among other configurations, can also be managed by an admin user or set as default for new users.
A user participating in certain studies is automatically granted access to corresponding content and modules.
For example, one or more questionnaires can be assigned to a study.
Depending on the use case, this way, users can decide what studies they want to take part in, which may grant or revoke access to corresponding modules.

The \emph{Questionnaire-Module} can represent multi-page questionnaires with common elements such as single/multiple-choice, text-answers, date-questions, and selection boxes or texts in various ways and any arrangement as displayed in Figure \ref{fig:corona-health-stress-questionnaire}.
Filled-out questionnaires are sent to the server as JSON answer sheets, where they can be evaluated in combination with the \emph{Feedback-Module} by configurable rules to generate feedback blocks accordingly.
Meta and sensor data can also be uploaded via these answer sheets.
The questionnaires also provide configurable notification schedules, which can remind the user to fill out a specific questionnaire.
In CC, the Feedback-Module is used as a COVID-19 symptom checker,
providing a first estimation based on the given answers.
Because at the time of planning the application (March 2020), the information for COVID-19 regularly updated, it was assured by design that adjustments to contents of questionnaires, feedbacks, and rules used to calculate the latter, as well as other extensive and critical changes, must be possible during live operation.
Also, system configuration adaptions must not lead to data inconsistencies or even wrong information for the users, especially in edge cases, where the user is running the app during the patch-phase either online or offline.
In order to properly test this, in addition to the complex synchronization logic, a test version of the system ran in parallel, in which the migration could be simulated.

\emph{Information} can be presented in many ways.
As a configurable format that works on both Android and iOS, HTML files are used for static content such as the privacy policy or legal information.
\emph{News} can be presented via Markdown files, which are easy-to-edit for the editors.
With this module, we keep the user up-to-date on current topics and regulations.
\emph{Tips} and \emph{FAQ} are also supported by the client framework.
Here, general information about the COVID-19 pandemic and information about protective behaviors and measures are offered.
Users can rate or like content, which, in turn, gives the authors feedback on its usefulness.

\emph{Offline availability:}
If a user is offline, e.g., when no internet connection is available, and gets a notification from the application, the corresponding questionnaire must be available anyway.
For non-critical parts of the application, contents can be stored locally.
This does not apply to \emph{Tips} and \emph{News}, because they must comply to the latest state of the pandemic's information and regulations, which are updated frequently.
Furthermore, feedback that may contain an individual recommendation is always loaded from the server on which the current rules are defined.

\subsection{User Privacy}
\label{sec:userPrivacy}
As the data collected by the platform is highly sensitive, protection of the acquired information is of utmost importance.

\subsubsection{Privacy by Design}
\label{sec:privacyByDesign}

One important aspect is, that the user does not login with any personal credentials.
For the sake of user privacy in the context of COVID-19, we have adapted and enhanced the manual registration and login process on the client-side with an automated, anonymously generated account.
In this way, the user is relieved from efforts associated with creating a user account and disclosing personal data, which is not even necessary for the purpose of these applications.
In consequence, data is only stored anonymously and the user does not have to remember any login data.
Through the automatically generated ID, we are still able to correlate connected data points from the same user.

To ensure that collected data do not identify the user, they are anonymized even before they are stored locally or sent to the backend.
The location, for example, is only recorded to an accuracy of 11.1 km, which does not identify the user, but is still sufficient for geographic evaluations.
With the explicit permission of the user, in the Android version, we can track coarse, aggregated app usage statistics via \emph{UsageStats}.

\subsubsection{User Awareness}
\label{sec:userAwareness}
User data protection and privacy are very sensitive topics in the field of EMA and MCS in general, but especially COVID-19 related ones, which often cause concerns by their users, as discussed in \cite{borra2020COVID19AppsPrivacySecurityConcerns}.
To counteract this, we
provide users with the best possible information about data usage and let them decide what they want to share in an anonymized way.
We present exactly what data is being collected at what point in time.
By design, the user stays anonymous and has the ability to actively opt-in to the collection of certain data.
This is done either by the OS's permission interface for location data and notification settings, or app-internal.
App-internal consent screens are shown before entering the main screen and creating a database entry, and before filling out questionnaires.
The data collection only takes place when the user actively submits a questionnaire by clicking the \emph{Submit}-Button, hence there is no background data collection.
Our privacy policy can be viewed at any time in the \emph{About Us} section of the application and on the corresponding website.

\begin{figure}
	\centering
	\begin{subfigure}[b]{0.15\textwidth}
		\centering
		\includegraphics[trim=3cm 0 3cm 0, clip=true, width=\textwidth]{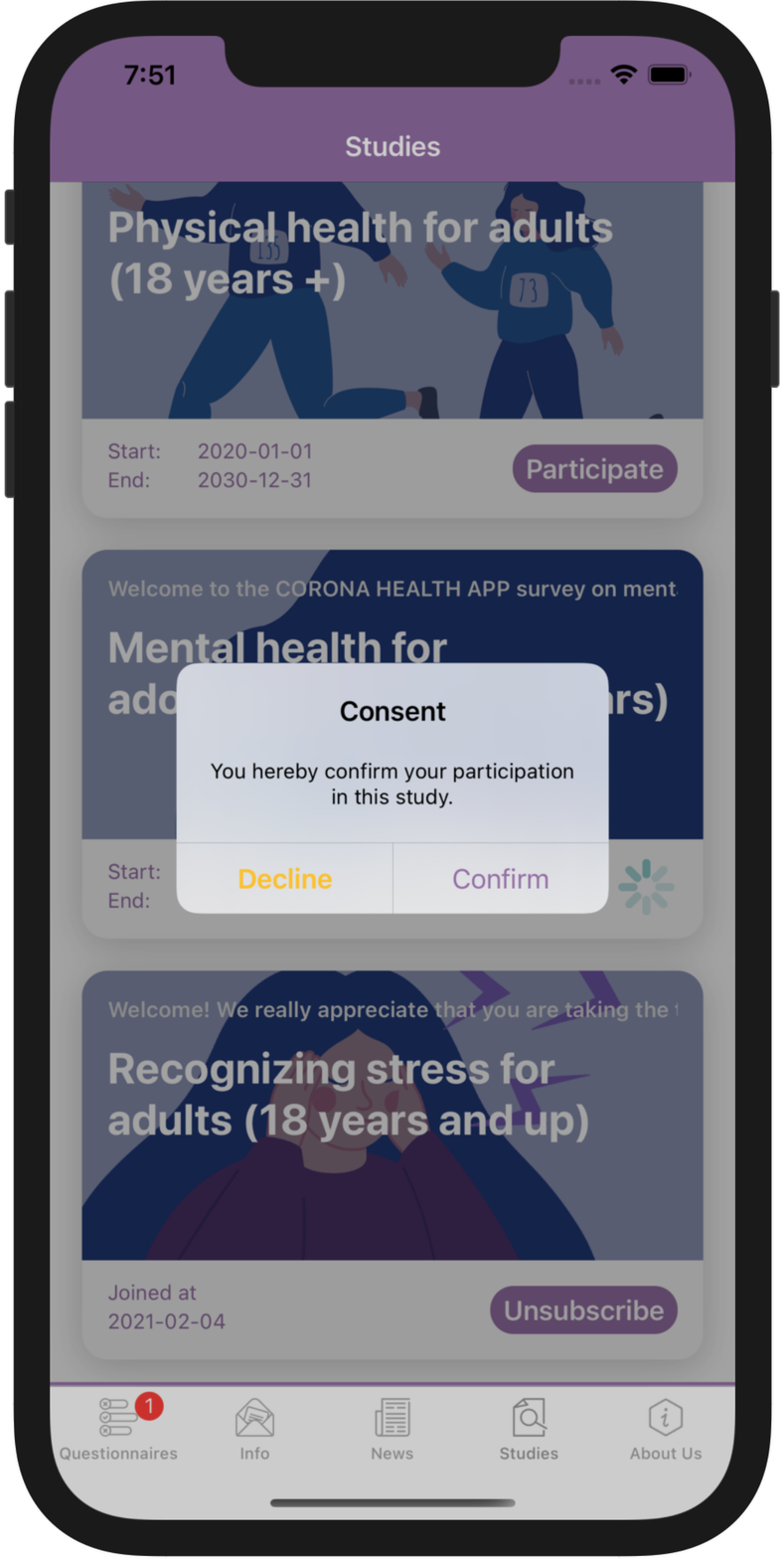}
		\caption{Studies}
		\label{fig:corona-health-stress-study}
	\end{subfigure}
	\hfill
	\begin{subfigure}[b]{0.15\textwidth}
		\centering
		\includegraphics[trim=3cm 0 3cm 0, clip=true,width=\textwidth]{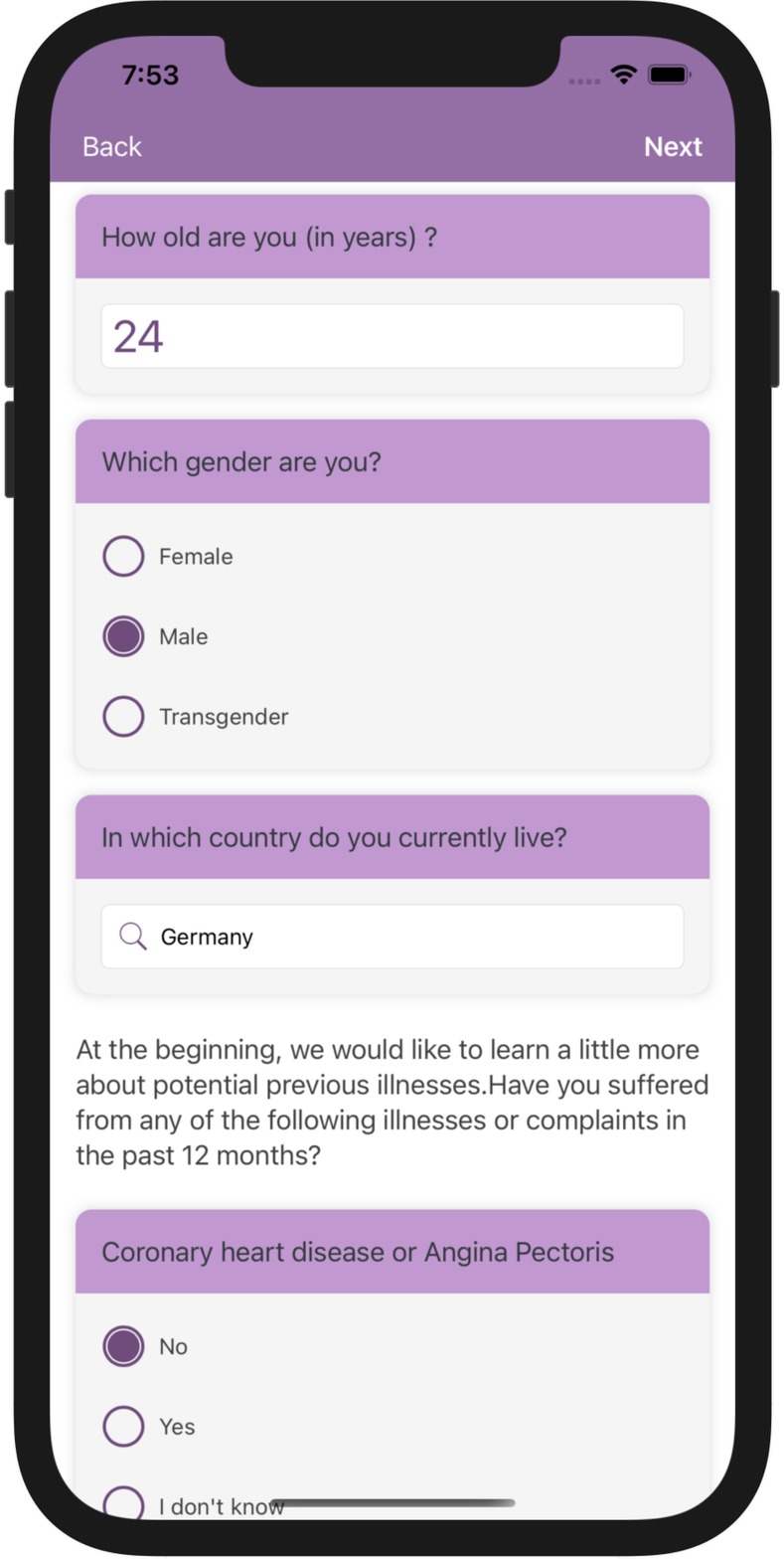}
		\caption{Questionnaire}
		\label{fig:corona-health-stress-questionnaire}
	\end{subfigure}
	\hfill
	\begin{subfigure}[b]{0.15\textwidth}
	\centering
	\includegraphics[trim=3cm 0 3cm 0, clip=true,width=\textwidth]{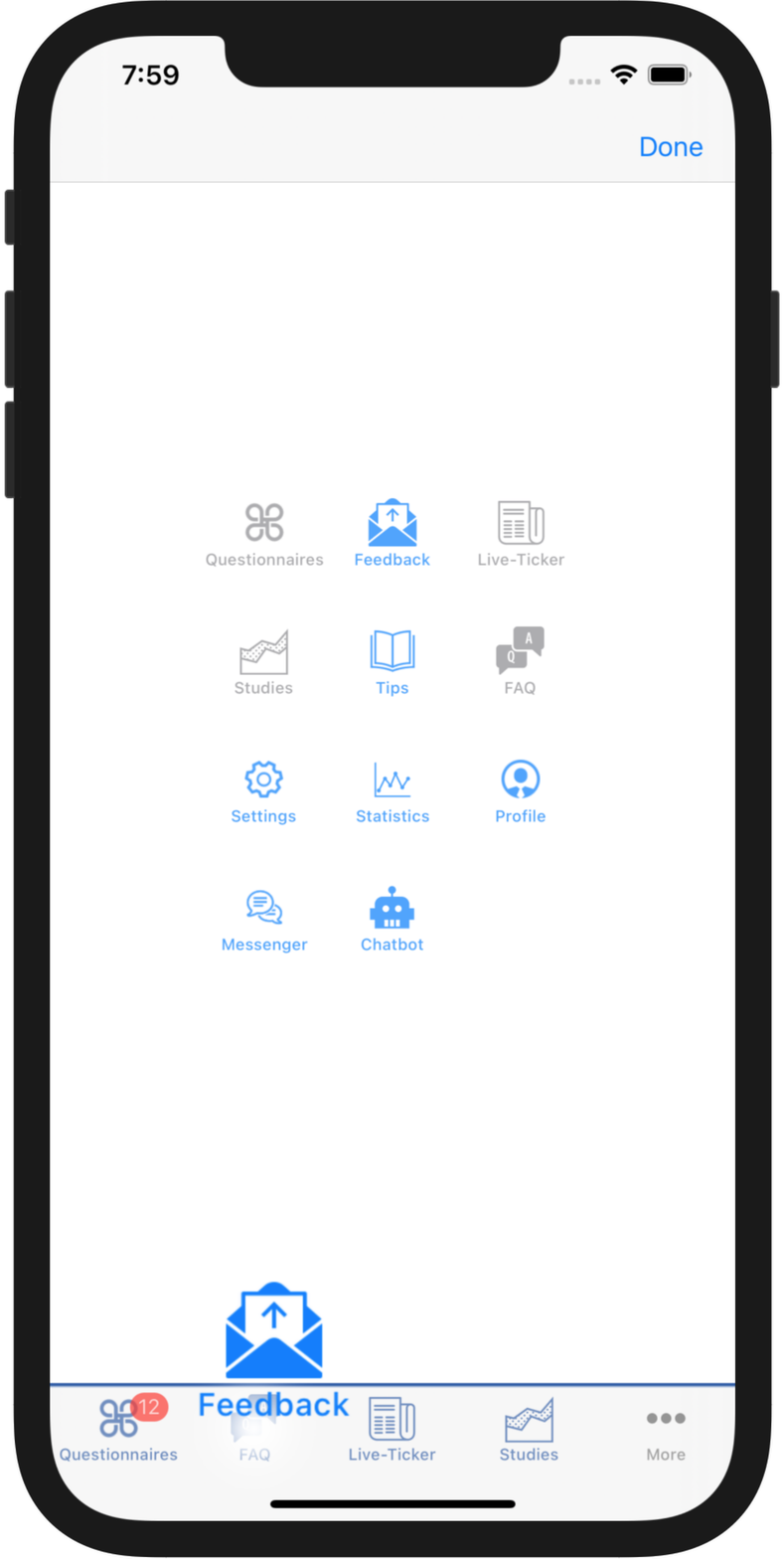}
	\caption{App configurator}
	\end{subfigure}
	\caption{iOS in-app screenshots.}
	\label{fig:track-your-health-configurator}
\end{figure}

\section{Timeline}
\label{sec:timeline}
\begin{figure}[htbp]
	\centerline{\includegraphics[trim=0 4cm 0 4cm, clip=true, width=1.0\linewidth]{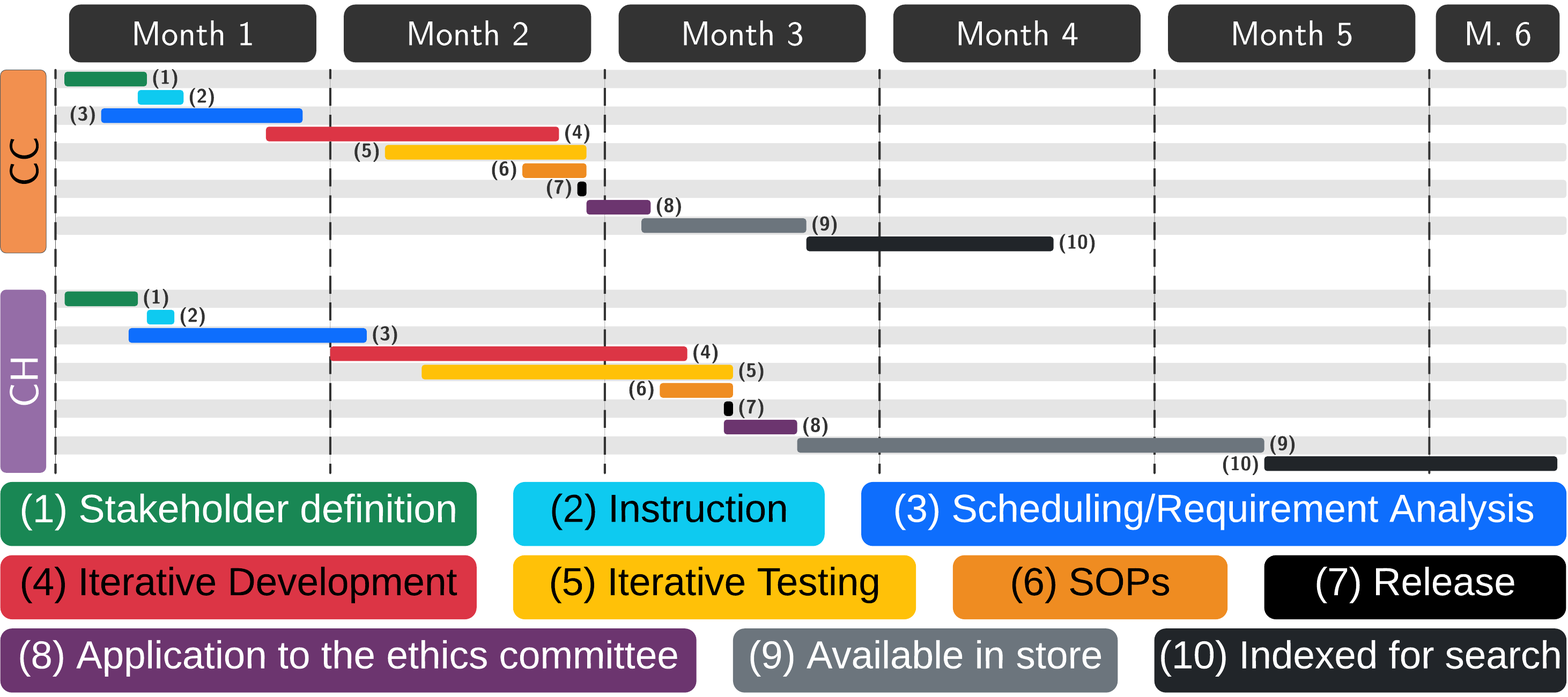}}
	\caption{Development time sequence milestones.}
	\label{fig:timeSequence}
\end{figure}

In this section, we
highlight some critical aspects outside the core software development
that posed challenges in releasing CC and CH.
Due to compliance with several regulations, the development process mostly followed the standard waterfall model.
Figure \ref{fig:timeSequence} schematically shows a Gantt-Chart of the development and release of CC and CH.

\subsection{Planning Phase}
During the planning phase, we defined the requirements for the COVID-19 apps together with the other stakeholders.
The main challenge here was to map expectations onto the feature set we already developed and to estimate the efforts to implement new modules of features.
In most cases, with small modifications to the requirements or extensions to our platform, we could keep the development effort low.
Solely, the \emph{News} module had to be implemented from scratch.
Here, we have provided the editors with a platform on which they can write and upload
COVID-19-related news articles in Markdown format.
In order to prevent the publication of misleading or incorrect information, news articles are only published if at least three experts have approved them.

For planning and live demonstration purposes, a configurable iOS demo version allowed an application to be quickly assembled from modules.
Figure \ref{fig:track-your-health-configurator} (c) shows the configurator's drag and drop menu for the different modules.
Here it is also possible to display demo data and interact with the latter.

\subsection{Development and Testing Phases}
During the development phase, we iteratively held consultations with medical experts and also provided assistance with the technical specification of the questionnaires and evaluation rules.
The test phase of beta-versions began slightly offset from these meetings.
Due to the ever changing COVID-19 situation, the requirements were adapted and therefore, the configuration files had to be updated correspondingly.
At this point in time, some steps were not fully automated and integrated in the grown platform.
Therefore, reconfiguring the live system, in combination with the translation process -- for up to eight languages per app -- required some manual work.

\subsection{Policy Compliance}
\label{sec:policyCompliance}
The entire system has been validated/verified on the basis of the IEC 62304 and IEC 82304 standards (medical device software / healthcare apps) as well as the GAMP 5 regulations (standard work of the pharmaceutical industry).
This comes with a huge effort in creating documentation and enforcing security and safety measures on the platform itself.
After completing the certification, changes to the platform are associated with a lot of post-processing of the documentation and re-testing, to restore the certification status.
The documents resulted in approx. 300 pages for CC and 860 pages for CH.

\subsection{Ethics Committee and Store Review Process}
\label{sec:ethicsCommitteeAndStoreReviewProcess}
\ifx\doubleBlind\undefined
With the completion of the certifications and the approval of our apps by the Ethics Committee of the University of Würzburg, we handed over the applications to the respective store review process.
\else
With the completion of the certifications and the approval of our apps by the Ethics Committee of \anonymize{XX xxxxxxxx X xxxxxxxx}, we handed over the applications to the respective store review process.
\fi
Due to the amount of misinformation about the COVID-19 and the potentially high damage that could be caused by malicious apps in this area, the procedures for releasing such apps are particularly strict.
\ifx\doubleBlind\undefined
This led to longer verification and indexing processes than we were used to from our other apps such as TrackYourTinntius \cite{probst2016EmotionalStatesMediatorsTinnitusLoudness}.
\else
This led to longer verification and indexing processes than we were used to from our other apps such as \anonymize{xxxxxxxxxxxxxxx}\cite{blindedSource1}.
\fi

\section{Conclusion and Future Work}
\label{sec:conclusionAndFutureWork}
In this paper, we presented TrackYourHealth (TYH), a generic, configurable EMA and MCS platform.
We introduced its core features and modules, and pointed out
how we implemented user privacy and data anonymization.
\ifx\doubleBlind\undefined
With TYH, we developed and released two multi-platform COVID-19-related apps, Corona Check and Corona Health, within a matter of weeks.
\else
With TYH, we developed and released two multi-platform COVID-19-related apps, \anonymize{xxxxxxxxxxxx} and \anonymize{xxxxx xxxxxx}, within a matter of weeks.
\fi
We presented the timeline for developing these apps
and highlighted the specific challenges
in the processes, including the large efforts necessary for
policy compliance with medical regulations.

For future work, we plan to add end-user programming to improve configuration and content management.
The goal is to streamline and automate parts of the process in which external partners are involved, especially from other disciplines.

\section*{Acknowledgment}
\label{sec:acknowledgments}
\ifx\doubleBlind\undefined
We are grateful for the support from Marc Holfelder\footnote{LA2 GmbH: \url{https://www.la2.de}; last accessed: 2021-02-22} during the certification process and for the software development efforts by Julian and Fabian Haug for the Android applications.
\else
We are grateful for the support from \anonymize{xxxxxxx xxxxx}\footnote{\anonymize{xxxxxxxxx: xxxxxxxxxxxxxxx}; last accessed: 2021-02-22} during the certification process and for the software development efforts by \anonymize{xxxxxxxxxxxxxxxxxxx} for the Android applications.
\fi

\bibliographystyle{IEEEtran}

\end{document}